\documentclass{elsart}
\usepackage{epsfig}
\usepackage{amssymb}

\usepackage{amsmath}

\begin{document}

\begin{frontmatter}

\title{Vortex ratchet}

\author{A. P\'erez-Madrid\corauthref{cor1}\thanksref{label1},}
\ead{agusti@ffn.ub.es}
\corauth[cor1]{Corresponding author}
\author{T. Alarc\'on\thanksref{label2},}
\author{J.M. Rub\'{\i}\thanksref{label1},}
\address[label1]{Departament de F\'{\i}sica Fonamental, Facultat de F\'{\i}sica,
Universitat de Barcelona, Avda. Diagonal 647, 08028 Barcelona, Spain}
\address[label2]{Centre for Mathematical Biology, Mathematical
Institute, University of Oxford, 24-29 St. Giles', Oxford OX1 3LB, United
Kingdom}

\begin{abstract}
We present a new class of thermal ratchets operating under the action of a
symmetry breaking non-Hermitian perturbation which rectifies thermal
fluctuations, and driven by a unbiased periodic force. The peculiar
non-Hermitian dynamics which follows causes energy transduction from the force
to the system in such a way that an average `uphill' particle current is
induced. We discuss physical realizations in assemblies of orientable particles,
in itinerant oscillator models, and in problems of diffusion in disordered
media \end{abstract}

\begin{keyword}
Brownian motor\sep transport processes\sep dissipation

\PACS  05.40.-a\sep 82.70.-y\sep 47.27.-i

\end{keyword}

\end{frontmatter}

% main text

\section{Introduction}
In the last few years it has been shown the existence a wide variety of
transport processes at the mesoscopic level in which thermal noise plays a
decisive role. To understand how those processes work, several
physical models and technological implementations have been proposed
\cite{reimann}, \cite{reimann2}. The peculiar effect of thermal noise can be
illustrated in thermal ratchets or Brownian motors, which have the ability of
extracting work from out-of-equilibrium fluctuations in spatially periodic
systems without spatial inversion symmetry. One way to do this is by the
combination of a ratchet-like potential which rectifies thermal fluctuations,
and a periodic unbiased force driving the system out of equilibrium.

Our purpose in this paper is to propose a new class of Brownian motors able to
extract work from thermal fluctuations in nonequilibrium systems driven by a
periodic force. Unlike the ones previously introduced, the symmetry breaking
is due to the presence of a vortex field responsible for the existence of a
non-Hermitian component in the stochastic dynamics. Fluctuations are then
rectified by a non-equilibrium source instead of a ratchet-like potential. We will call
those devices \emph {vortex ratchets}.

The paper is organized as follows. In section 2, we discuss the stochastic
dynamics of these systems. We formulate the Fokker-Planck equation
 and compute the susceptibility. In section 3, we analyze the
dissipation of energy in the system and introduce the ratchet current.
Section 4 is devoted to discuss some applications

\section{Stochastic dynamics}

We consider a Brownian degree of freedom, parameterized by a coordinate $%
\mathbf{x}$, interacting with a thermal bath which is maintained out of
equilibrium by the persisting action of an external drift $\mathbf{v}(%
\mathbf{x})$, \cite{alarcon}. This drift could represent, for example, a
constant or a quenched velocity field or an external field. The stochastic
dynamics is governed by the probability density $\Psi (\mathbf{x},t)$ satisfying
the conservation law

\begin{equation}  \label{eq:f-p2}
\partial_t\Psi(\mathbf{x},t)+\nabla_{\mathbf{x}}\cdot(\mathbf{v}(\mathbf{x}%
)\Psi(\mathbf{x},t)) =-\nabla_{\mathbf{x}}\cdot\mathbf{J}_{\psi}(\mathbf{x}%
,t)\; ,
\end{equation}

\noindent in which the probability current $\mathbf{J}_{\psi}$ is given by

\begin{equation}  \label{eq:current}
\mathbf{J}_{\psi}(\mathbf{x},t)=-D \nabla_{\mathbf{x}}\Psi(\mathbf{x},t)+ b%
\mathbf{F}(\mathbf{x},t)\Psi(\mathbf{x},t)\; .
\end{equation}

\noindent The dynamics of the Brownian degree of freedom is then governed by
the Fokker-Planck equation

\begin{equation}  \label{eq:f-p}
\partial_t\Psi(\mathbf{x},t)=-\nabla_{\mathbf{x}}\cdot(\mathbf{v}(\mathbf{x}%
)\Psi-D \nabla_{\mathbf{x}}\Psi)-\nabla_{\mathbf{x}}\cdot b\mathbf{F}(%
\mathbf{x},t)\Psi(\mathbf{x},t)\; ,
\end{equation}

\noindent where $b$ is the mobility, $D=k_BTb$ the corresponding diffusion
coefficient, and $\mathbf{F}(\mathbf{x},t)=\mathbf{F}_o(\mathbf{x}%
)\lambda(t) $ a periodic force, with $\lambda(t)=\lambda_o\text{e}^{i\omega
t}$.

In the linear response regime the formal solution of the Fokker-Planck
equation reads

\begin{equation}  \label{eq:solution}
\Psi(\mathbf{x},t)=\Psi_o(\mathbf{x})+\int_{t_o}^t \lambda(t^{\prime})\text{e%
}^{(t-t^{\prime})\mathcal{L}_o}\mathcal{L}_1\Psi_o(\mathbf{x}%
)dt^{\prime}=\Psi_o(\mathbf{x})+\triangle\Psi(\mathbf{x},t)\, ,
\end{equation}

\noindent where $\mathcal{L}_o=-\nabla_{\mathbf{x}} \cdot\mathbf{v}+D\nabla_{%
\mathbf{x}}^2$ is the unperturbed Fokker-Planck operator, and $\mathcal{L}%
_1=-b\nabla_{\mathbf{x}}\cdot\mathbf{F}_o$ the perturbation. Moreover, $\Psi_o(%
\mathbf{x})$ corresponds to the stationary solution of eq. (\ref{eq:f-p}),
\cite{hanggi-tomas}.

The presence of the perturbation causes deviation in the coordinate, $\Delta 
\mathbf{x}$. To compute that quantity, we will expand the term $\mathcal{L}%
_1\Psi_o(\mathbf{x})$ in series of the eigenfunctions of the operator $%
\mathcal{L}_o$, $\phi_n$ with eigenvalues $a_n$; $n=0,1,....$

\begin{equation}  \label{eq:expansion}
\mathcal{L}_1\Psi_o(\mathbf{x})=\sum_{n=0}^\infty \{c_n\phi_n(\mathbf{x}%
)+c_n^\ast\phi_n^\ast(\mathbf{x})\}\; ,
\end{equation}

\noindent where $c_n$ are the corresponding coefficients. We obtain

\begin{equation}  \label{eq:respuesta}
\Delta \mathbf{x} (t)\,=\,\int \,\mathbf{x} \Delta \Psi (\mathbf{x} ,t)d%
\mathbf{x} \,=\,\int_{t_o}^{t}d\tau {\boldsymbol {\chi}} (t-\tau )\lambda
(\tau ),
\end{equation}

\noindent which defines the susceptibility ${\boldsymbol{\chi}} (t)$.

We will assume the existence of a dominant time scale governing the
relaxation process, corresponding to the $n=1$ mode in the expansion (\ref%
{eq:expansion}). Since the remaining modes decay faster, we can truncate the
series retaining only the second term. Thus, considering only contributions
of the first mode the susceptibility is given by ${\boldsymbol{\chi}}
(t)\,=\,\mathbf{A}e^{a_{1}t}+c.c.$, with $\mathbf{A}$ defined as

\begin{equation}  \label{eq:A}
\mathbf{A}=c_{1}\int \mathbf{x} \phi _{1}(\mathbf{x} )d\mathbf{x}
\end{equation}

\noindent Assuming now $t_o\rightarrow -\infty$, eq. (\ref{eq:respuesta})
can be rewritten as

\begin{equation}  \label{eq:respuesta2}
\Delta\mathbf{x} (t) ={\boldsymbol {\chi}}(\omega)\lambda(t)\; ,
\end{equation}
\noindent where ${\boldsymbol {\chi}} (\omega)$ is the Fourier transform of $%
{\boldsymbol {\chi}} (t)$ given by

\begin{equation}  \label{eq:susceptibility}
{\boldsymbol {\chi}}(\omega )\,=\frac{\mathbf{A}}{I_{1}}\frac{1}{\beta
-i(\alpha +1)}+ \frac{\mathbf{A}^{*}}{I_{1}}\,\frac{1}{\beta-i(\alpha -1)}\;
,
\end{equation}

\noindent with $^{*}$ standing for complex conjugate. Due to the
non-Hermitian nature of the operator, the first eigenmode is complex:
$a_{1}\equiv R_{1}+iI_{1}$ with $R_{1}$ and $I_{1}$ being its real and imaginary
parts, respectively. The remaining parameters in eq. (\ref{eq:susceptibility})
are $\beta \equiv R_{1}/I_{1}$ and the normalized frequency $\alpha \equiv
\omega /I_{1}$.

\begin{figure}[htb]
\centerline{\psfig{file=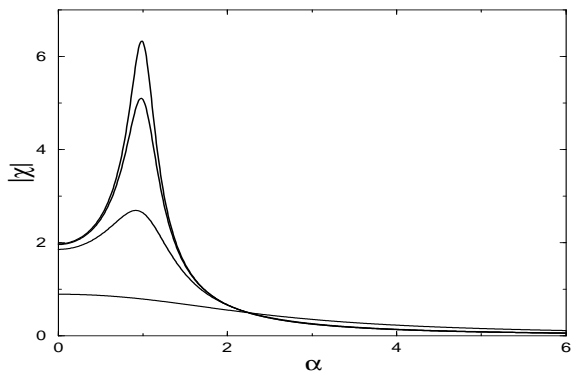,height=5cm,width=7.5cm}}
\caption{Non-dimensional modulus of the susceptibility
as a function of $\alpha$, for different values of the parameter
$\beta$, the smaller the value of $\beta$ the sharper the curve. The
resonance fades away practically for $\beta\approx 10$.} \label{fig:1}
\end{figure}

In Fig. 1, we show that during the relaxation process of non-equilibrium
fluctuations the susceptibility undergoes a resonant
behavior when the frequency of the force matches the imaginary part of the
first eigenvalue of the non-perturbed operator $\mathcal{L}_o$. This
behavior reveals the existence of a resonant coupling between the periodic
force and the non-equilibrium source, responsible for the non-Hermitian
nature of $\mathcal{L}_o$.The implications of that coupling in the energy
transduccion of the system will be analyzed in the next section.

\section{Ratchet effect}

Systems governed by the non-Hermitian dynamics discussed in the previous
section may transduce the energy supplied by an unbiased periodic force into
kinetic energy, thus inducing a net particle current. To analyze this peculiar
behaviour, we will first calculate the power dissipated by the system. To
that purpose we will apply the scheme of mesoscopic non-equilibrium
thermodynamics \cite{agusti}. For a system described by the
probability density $\Psi (\mathbf{x},t)$, the variation of the entropy due
to changes in configurations in $\mathbf{x}$-space is given by

\begin{equation}  \label{eq:entropyvariation}
\delta S=-\frac{1}{T}\int\mu\delta\Psi d\mathbf{x\, ,}
\end{equation}

\noindent where $\mu=k_BT\text{ln}\Psi+U$ is the chemical potential, $U(%
\mathbf{x},t)$ the potential, and T the temperature. The rate of change of the
entropy can be obtained by taking the time derivative of eq.
(\ref{eq:entropyvariation}), and using eq. (\ref{eq:f-p2}). One achieves

\begin{equation}  \label{eq:rateentropy}
\frac{dS}{dt}+\int\psi\mathbf{v}\cdot\nabla(\mu/T)d\mathbf{x}=- \int\mathbf{J%
}_{\psi}\cdot\nabla(\mu/T)d\mathbf{x}\; ,
\end{equation}

\noindent The right hand side of eq. (\ref{eq:rateentropy}) constitutes the
irreversible part of the rate of change of the entropy or entropy
production. Consequently, the power supplied by the external force and dissipated into the system is
obtained from eq. (\ref{eq:rateentropy}) by using the expression of the chemical
potential. One obtains

\begin{equation}  \label{eq:externalpower}
P_F=\int\mathbf{J}_{\psi}\cdot\mathbf{F}(\mathbf{x},t) d\mathbf{x}=
\mathbf{F}_o\cdot\left\{\frac{d}{dt}\langle \mathbf{x}\rangle-\langle
\mathbf{v}(\mathbf{x})\rangle\right\}\lambda(t)\; ,
\end{equation}

\noindent which defines the particle current $\langle \dot{\mathbf{x}}%
\rangle=d\langle \mathbf{x}\rangle/dt-\langle \mathbf{v}(\mathbf{x})\rangle$%
. To obtain eq. (\ref{eq:externalpower}), we have assumed an homogeneous force
and used eqs. (\ref{eq:f-p2}) and (\ref {eq:current}). Thus, $P_F$ can be
interpreted as the projection of the particle current along the direction of the
oscillating force.The quantity of interest in experiments is the time-averaged
dissipated power, defined as

\begin{equation}  \label{eq:averagepower}
P(\omega)=\frac{\omega}{2\pi}\int_0^{2\pi/\omega}P_F dt\; ,
\end{equation}

\noindent This quantity is not only a function of the imaginary part of the
susceptibility as occurs in Hermitian systems. The presence of the external
drift introduces a more complicated dependence on the moments of $\mathbf{x}$.
Moreover, it does not have a definite sign, and in general can be a
non-monotonous function of the frequency expressing the resonant character of
the energy dissipation. Since in
general the external drift $\mathbf{v}(\mathbf{x})$ introduces a characteristic
frequency in the system playing the same role  of a vorticity, this systems
behaves as a vortex ratchet.

\section{Applications}

Our purpose in this section is to present different manifestations of the
vortex ratchet, as well as to indicate
potential applications in different fields.

\subsection{Orientable particles}

An orientable particle of mesoscopic size in a vortex field under the influence
of a periodic force exhibits the phenomenology discussed previously. In absence
of the external force the orientation of the particle is characterized by a
director vector undergoing Brownian motion which on average rotates at the
velocity imposed by the local vorticity, $\boldsymbol{\omega}_{o}$. The most
interesting situation occurs when the force is perpendicular to the vortex
field. In that case, if $\omega >1/2\omega _{o}$ the particle
acquires an excess of angular velocity as a consequence of the torque it feels.
In such a situation,  the energy supplied by the periodic force is transduced
into rotational kinetic energy, thus inducing a net particle current. This power
is given by $P_F=\boldsymbol{\tau}\cdot(\boldsymbol{\Omega}_P-\frac{1}{2}
\boldsymbol{\omega}_o )$, where $\boldsymbol{\Omega}_P$ is the average angular
velocity of the particle,and $\boldsymbol{\tau}$ the torque acting on it. In
this range of frequencies the system behaves as a Brownian motor.

The director could represent a dipole moment oriented by a field.
Suspensions of such dipoles in a liquid phase exhibit peculiar collective
behaviours. This is the case of electro- and magneto-rheological fluids,
ferrofluids \cite{bacri}, \cite{agusti2}, \cite{reimann3} and dilute solution of
rod-like polymers \cite{doi}.

\subsection{Itinerant oscillator models}

The itinerant oscillator model  essentially consists of a Brownian particle
with an orientable core coupled via an interaction potential to the shell. It
was proposed to explain microwave
dielectric absorption of polar fluids\cite{coffey}. A particular realization
of the model is an inhomogeneous body under the influence of a vortex field and
a constant force perpendicular to it. The inhomogeneity induces a
time-dependent dipole moment. The associated torque is
$m(t)\mathbf{g}\times\mathbf{r}$, with $m(t)$ being the dipole
moment strength, $\mathbf{g}$ the constant external force, and $\mathbf{r}$ the
orientation vector. When the dipole moment varies periodically in time, as in
the case of the orientable particle, the system behaves as a vortex ratchet.
Unlike the previous case, variations in the exerted torque are due to internal
reorganizations and not to variations of the external field. The itinerant
oscillator model may mimic a living cell with an inhomogeneous density
distribution in an external field, a particle in a cage formed by other
particles, a rod-like polymer moving in a tube \cite{doi}, or a monodomain
magnetic particle whose magnetic moment undergoes fluctuations. Those systems
would be good candidates to act as vortex ratchets.

\subsection{Diffusion through random and structured media}

A particle diffusing in a randon media advected by a steady
mean-flow velocity $\mathbf{v}_{o}(\mathbf{x})$ in the presence of a
periodic force also manifests the ratchet effect previously discussed. The
dynamics of the particle follows from the Fokker-Planck equation
(\ref{eq:f-p} ), where the drift $\mathbf{v}(\mathbf{x})$ is now a random
velocity distributed around the mean-flow velocity according to a Gaussian
probability distribution with variance $\Gamma$. The Fourier transform of the
random drift $\mathbf{v}(\mathbf{k})$ displays both longitudinal and transversal
correlations

\begin{equation}\label{eq:correlations}
\langle v_i(\mathbf{k})v_j(\mathbf{k}')\rangle = 2(2\pi)^3 \Gamma
\delta_{ij}\delta(\mathbf{k}+\mathbf{k}')\; .
\end{equation}

\noindent The transversal correlations
play the role of a vorticity, introducing the non-Hermiticity in the dynamics.

Assuming that the external force is homogeneous, the
dissipation is obtained through eqs. (\ref{eq:externalpower}) and
(\ref{eq:averagepower}), after averaging over the disorder. In Fig. 2,
we have represented the normalized dissipated power  $P$ as a function of the
normalized frequency $\alpha$, for two levels of disorder. These levels are
characterized by the parameter $s\equiv\sigma^2(Dk^2)^2$
where $D$ is the diffusion
coefficient and $\sigma^2\equiv\Omega/2k^2\Gamma$, with $\Omega$ the volume of
the system. The disorder then increases when decreasing $s$.

\begin{figure}[htb]
\vspace{0.5cm}
\centerline{\psfig{file=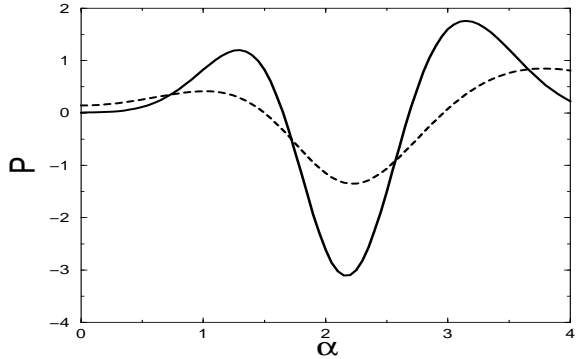,height=5cm,width=7.5cm}}
\caption{Contribution of the external force to the dissipated energy as a function
of $\alpha$. Solid line corresponds to s=2, dashed line represents the same quantity for s=1.}
\label{fig:2}
\end{figure}

The figure shows how the contribution of the periodic force to the total
dissipated power exhibits a minimum when the frequency matches the
characteristic frequency of the system, thus revealing the resonant character of
the dissipation. Due to the characteristics of this
system, when the power is negative the particle current is positive. The figure
also shows that the power or, in view of eq. (\ref{eq:externalpower}),
the current is positive for frequencies around the resonance frequency,
which manifests that the velocity of the particles is larger than the average
drift. Therefore, in those conditions the system acts as a Brownian motor.

A similar phenomenon  occurs in a spatially periodic two-dimensional pattern of
triangular vortexes when a two-dimensional  oscillating force
is applied, at sufficiently high Reynolds number. In this conditions, a
large-scale current appears. This effect is accompanied by a reduction of the
dissipation in the system due to the induction of a negative eddy
viscosity\cite{sivashinsky}.

\section*{Acknowledgements}
This work has been supported by DGICYT of the Spanish Government under Grant No.
BFM 2002-01267.

\end{document}